# Apollo Video Photogrammetry Estimation
# Of Plume Impingement Effects

Christopher Immer
*ASRC Aerospace, M/S ASRC-15, Kennedy Space Center, FL  32899*
christopher.immer@ksc.nasa.gov
John Lane
*ASRC Aerospace, M/S, ASRC-15, Kennedy Space Center, FL  32899*
john.lane@ksc.nasa.gov
Philip Metzger
*NASA/KSC Granular Mechanics and Surface Systems Laboratory, KT-D3, Kennedy Space Center, FL 32899*
Philip.T.Metzger@nasa.gov
Sandra Clements
*ASRC Aerospace, M/S ASRC-15, Kennedy Space Center, FL  32899*
sandra.clements@ksc.nasa.gov

**Abstract**

The Constellation Project's planned return to the moon requires numerous landings at the same site. Since the top few centimeters are loosely packed regolith, plume impingement from the Lander ejects the granular material at high velocities. Much work is needed to understand the physics of plume impingement during landing in order to protect hardware surrounding the landing sites. While mostly qualitative in nature, the Apollo Lunar Module landing videos can provide a wealth of quantitative information using modern photogrammetry techniques. The authors have used the digitized videos to quantify plume impingement effects of the landing exhaust on the lunar surface. The dust ejection angle from the plume is estimated at 1-3 degrees. The lofted particle density is estimated at $10^8$-$10^{13}$ particles/m$^3$. Additionally, evidence for ejection of large 10-15 cm sized objects and a dependence of ejection angle on thrust are presented. Further work is ongoing to continue quantitative analysis of the landing videos.

## 1   Introduction

*1.1   Apollo Video*

For the six manned missions to the lunar surface during the Apollo program, video was recorded during the descent and landing. The videos contain an immense volume of qualitative information. Digitized versions of the film taken during the six lunar surface landings allow unprecedented analysis with modern photogrammetry techniques. While the intent of the videos was not necessarily designed with it in mind, many quantitative estimates of phenomena in the field of view of the landing cameras can be calculated.

Each of the six Apollo mission Landers touched down at unique sites on the lunar surface. Aside from the Apollo 12 landing site located 180 meters from the Surveyor III Lander, plume impingement effects on ground hardware during the landings were largely not an issue. With high vacuum conditions on the moon ($10^{-14}$ to $10^{-12}$ torr) (Heiken, Vaniman et al. 1991), motion of all particles is nearly ballistic. The surface of the Surveyor III craft was scoured and pitted by the ejected regolith.

The surface damage to Surveyor III permits estimation of the impact velocity of the exhaust ejecta (Katzan and Edwards 1991). Initial estimates from the shadowing of sand blasted surfaces were around 40 m/s for the particle velocity (Nickle and Carroll 1972). Further refinements bounded the minimum velocity of the particles to be greater than 70 m/s (Jaffe 1972) to 100 m/s (Cour-Palais, Flaherty et al. 1972). The most reliable estimate to date is based on the surface structure of the pitting, bounding the velocity in the range of 300 to 2000 m/s (Brownlee, Bucher et al. 1972). Note that the escape velocity from the lunar surface is around 2373 m/s.

*1.2 Constellation*

The Vision for Space Exploration has mandated that NASA return humans to the Moon and subsequently land people on the surface of Mars (Bush 2004). Contrary to the Apollo program, the Constellation Project's planned return to the moon requires numerous landings at the same site. Since the top few centimeters are loosely packed regolith, plume impingement from the Lander ejects the granular material at high velocities. Any hardware surrounding the landing sites would be subject to regolith impacts.

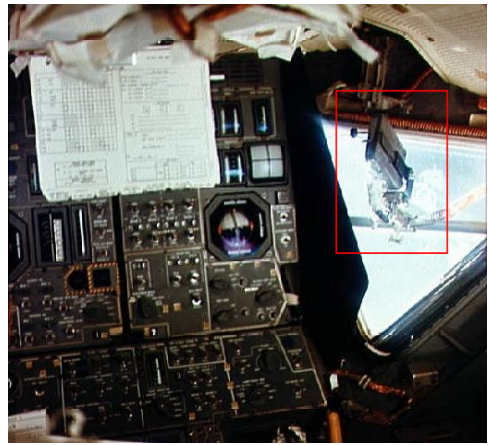

**Figure 1 Forward facing interior of the LM for Apollo 11. The sequence camera is positioned for viewing through the right window.**

There are two broad categories: dust impact and dust contamination (Wagner 2006). Dust impacts would result in damage to surfaces such as: pitting of optical windows, obscuration of viewports, damage to solar cells, damage to pressurized vessels, scouring of thermal blankets. These would be a direct result of regolith ejected during landing and will be the major focus of this paper.

Dust contamination would also result from transport of the lunar regolith. During the Apollo astronauts' short stays on the Moon, the lunar soil proved to be one of the most challenging obstacles. After several lunar excursions, with tenacious adhesion, dust coated everything: thermal radiators, space suits, viewports, environmental seals, the cabin interior, and even the astronauts themselves. The dust was such a problem on Apollo 12 that, during Pete Conrad's third extravehicular activity (EVA), the seals on his suit leaked at 0.25 psi/min, dangerously close to the safety criteria of 0.30 psi/min. An additional EVA would have been impossible on that mission (Gaier 2005).

In order to protect hardware, mitigation techniques must be developed. Several possibilities include: berm construction, shielding, stabilizing the landing area, and increasing distance. To develop mitigation techniques and to prove their effectiveness, it is imperative to understand the physics of plume impingement on the lunar surface.

*1.3 Sequence Camera*

The video analysis presented here for the descent to the lunar surface were all taken by the 16mm sequence camera. The camera was mounted in the lunar module cockpit facing out of the right side viewing window. Figure 1 shows an image of the interior of the lunar module cockpit. Typically, the camera had a 10mm lens and recorded at 6 frames per second. Digitized version of the film used for the analysis presented here were obtained from archives at Johnson Space Center.

**2  Results and Discussion**

*2.1  Camera Angle Measurements*

In order to extract quantitative information from the Apollo videos, it is necessary to know the angle of the camera with respect to the ground. Fortuitously, the lunar surface has been bombarded with meteors and many of the craters left behind are a perfect sub-fraction of a sphere. Assuming that the crater rim is perfectly circular, it is possible to calculate the camera view angles relative to the crater rim.

More specifically, by taking the ratio of the minor and major axes of the ellipse that is the 2D projection of the circular rim, the angle of the camera can be calculated:

$$\phi = \arcsin \frac{b}{a} \qquad (1)$$

where a and b are the major and minor axes of the ellipse of the crater as viewed through the camera and $\phi$ is the angle of the camera relative to the plane parallel to the plane of the crater rim.

Figure 2.2.1 shows the view during descent to the lunar surface of the Apollo 15 Lander. A crater can be seen in the upper left field of view with the major and minor axis shown. These measurements can be used to calculate the camera angle relative to the rim of the crater, which is approximately the angle of the camera relative to the lunar surface.

Proximity of a circular crater rim will be one of the criteria for scene selection for extraction of the dust angle. Fortunately, the method used to estimate the dust ejection angle only has weak dependence on absolute camera angle such that this method of camera angle estimation is adequate.

*2.2   Altitude Estimation*

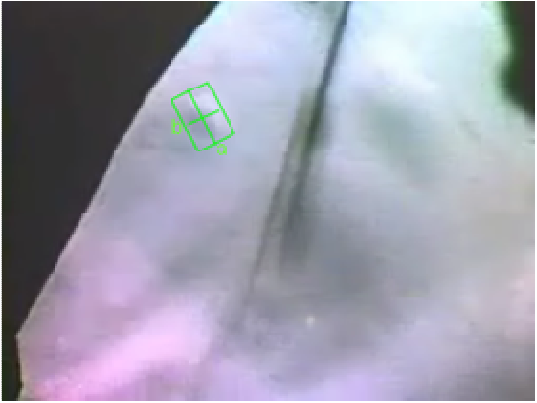

**Figure 2.2.1 View from window during descent to the lunar surface of Apollo 15. The crater major and minor axes (a, b) can be used to calculate the camera angle.**

The Apollo Lunar Modules were outfitted with 4 landing radars to measure the attitude and altitude during the landing procedure. This data would be extremely beneficial to quantify objects in the videos. The existing mission reports only have large scale plots of the altitude and attitude for the entire duration of the landing: they are insufficient resolution to be useful for the last 100 feet of landing. Theoretically the mission data tapes contain records of the descent trajectory, however these data are either completely lost or inaccessible.

In order to estimate the altitude of the Lander for scale calculation from the videos, the audio track was used. In it, the astronaut would call out the altitude, as read from the altimeter, every few seconds. The time and audio altitude were recorded for each of the missions. A rough descent trajectory can be calculated by linear interpolation between the audible points. Knowing the touch-down time as a common reference the estimated altitude fore each frame of the video can be calculated.

*2.3   Dust Ejection Angle By Shadow Elongation*

One of the primary goals of analyzing the Apollo landing videos was to extract the exhaust-plume induced dust ejection angle. In the videos, shortly before touchdown for 5 of the 6 landings, the shadow of at least one of the footpad/contact probes is visible. The footpad is a spherical section with the contact probe extending below it. Evident in the videos is a stretching or elongation of the footpad's shadow on the ground.

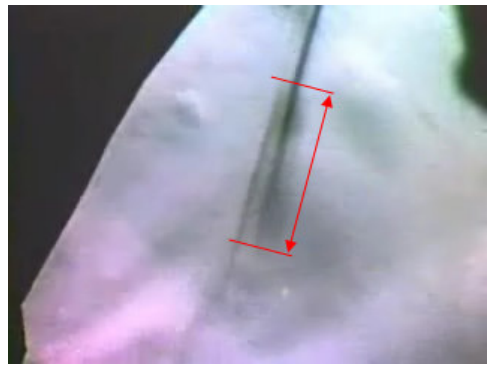

**Figure 2 Video from Apollo 15 landing just before touch down. Notice the elongation of the footpad's shadow.**

Figure 2 shows an example of the elongation witnessed during the landing of Apollo 15. The elongation occurs because there is a "sheet" of dust being eject radially from the point directly below the Lander engine. Since the sheet has a finite thickness and is partially transparent, the shadow elongation represents the projection of the shadow through the entire volume of the dust sheet. The side nearest to the Lander and faintest to see is the top of the dust sheet, while that furthest away is the sharp, distinct shadow on the lunar surface. Three techniques were employed to investigate shadow elongation as a means for dust angle ejection measurement.

### 2.3.1 Dust Ejection Angle Scale Model

A scale model of the Apollo lunar Lander was used along with a collimated light source to confirm the shadow elongation effect of the dust cloud and to help develop the geometrical/trigonometrical solution. A chamber was constructed of Plexiglas sheets, sealed on all sides with polyethylene, and the volume filled with smoke particles using an off-the-shelf fog machine. Using this model the geometry depicted in section 2.3.3 was confirmed. The top Plexiglas sheet could be positioned at various angles to model the top surface of the dust ejection angle. Figure 3 shows the shadow elongation of the scale model with the smoke filled Plexiglas chamber. Similar viewpoints to that seen from the right cockpit window are viewable both by removing the top portion of the model or by using a bore scope attachment to the camera to view from the window of the model.

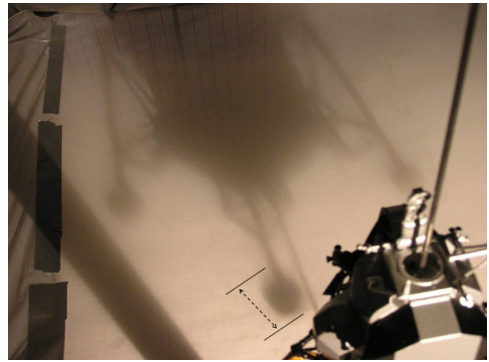

**Figure 3 Picture showing shadow elongation on scale model using a smoke-filled Plexiglas chamber.**

### 2.3.2 Dust Ejection Angle by 3D Cad

A high-fidelity Pro-Engineer (digital) model of the Apollo Lunar module was used to additionally help develop the geometry/trigonometry. The model was originally developed by Scott Sullivan and licensed by NASA for use to aid with the Constellation Program (Sullivan 2004). The virtual model of the scene adds much more flexibility, is more precise, and provides more rapid means for investigation of alternate geometries than the physical model. Additionally rendered video can be produced for different scenarios. Figure 4 shows some examples of modeling rendered in MAYA (a 3d Modeling/Visualization package).

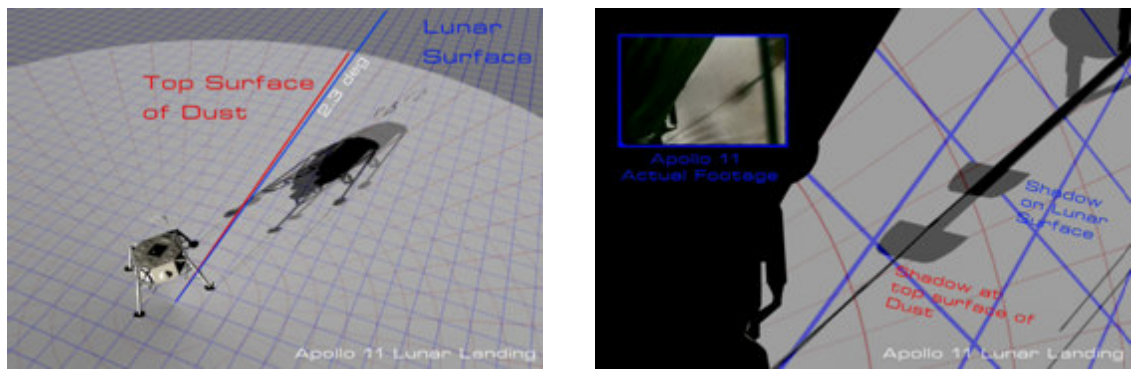

**Figure 4 Pro-Engineer/MAYA modeling of dust induced shadow elongation.**

## 2.3.3 Geometry/Trigonometry

Once confidence in the effect and the feasibility of the method from the scale model and computer model were in place, the analytical geometry/trigonometry was developed to calculate the dust ejection angle. Photogrammetry from the video and data from the mission reports were the primary sources for the unknowns used to calculate the ejection angle. The analytical calculations are summarized in Figure 5.

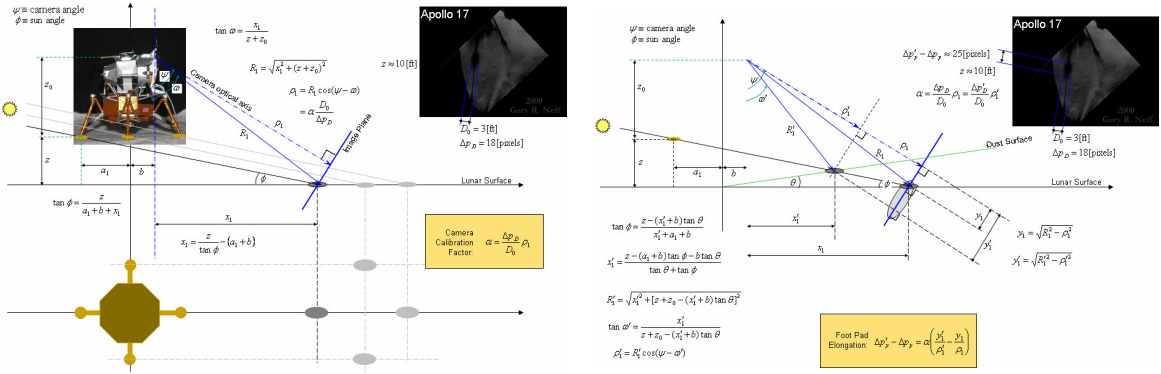

**Figure 5 Geometrical and trigonometrical relationships for the shadow elongation for the Apollo landing Videos. The equations in the right pane can be solved iteratively to match the elongation.**

For each of the landings the sun angle is known. The altitude of the Lander can be estimated by the technique detailed in section 2.2. The camera angle can be calculated as detailed in section 2.1. The video/shadow scaling from pixels to real units can be estimated by measuring the width of the footpad in pixels. With all these parameters the equations in the right pane of Figure 5 can be iterated, adjusting the dust ejection angle (the only unknown) to match the actual elongation in the video in pixels. A summary of the results averaged for all possible frames for each of the landing videos is shown in Table 1.

It was not possible to estimate the dust ejection angle for Apollo 12 because the sun angle was much lower than the other missions (about 5 degrees compared to about 11 for the other missions) and the footpad shadow enters the field of view just before landing (Orloff 2000). In general the ejection angle with respect to the surface is 1-3 degrees, however Apollo 15 landed on an inclined surface of about 11 degrees that caused a larger ejection angle. Additionally, immediately preceding touchdown, Apollo 15 had a "blow out" with very high dust ejection angle, likely greater than 22 degrees. This was most likely due to landing straddled over the rim of a crater.

| Apollo Mission | Dust Angle [deg] |
|---|---|
| 11 | 2.6 |
| 14 | 2.4 |
| 15 | 8.1 |
| 16 | 1.4 |
| 17 | 2.0 |
|  |  |
| avg | 3.3 |

**Table 1 Dust ejection angle for Apollo Missions derived from video photogrammetry.**

*2.4  Dust Loading From Luminosity*

From the landing videos, it is possible to obtain an estimate of the lofted density of dust ejected from the exhaust plume. The key to the estimation is that occasionally, for some sequential frames, the dust momentarily clears. For these sequential frames, if a bright object can be recognized in both the dusty and cleared frames then the normalized ratio of the bright object with and without the dust can be used to estimate the dust loading. For these calculations, the 8-bit digitized value of the frames were assumed to be linearly proportional to the light intensity. To normalize the light levels, a "shadowed" area with and without the dust can be measured. Figure 6 shows a summary of the calculations used to approximate the plume-induced dust loading.

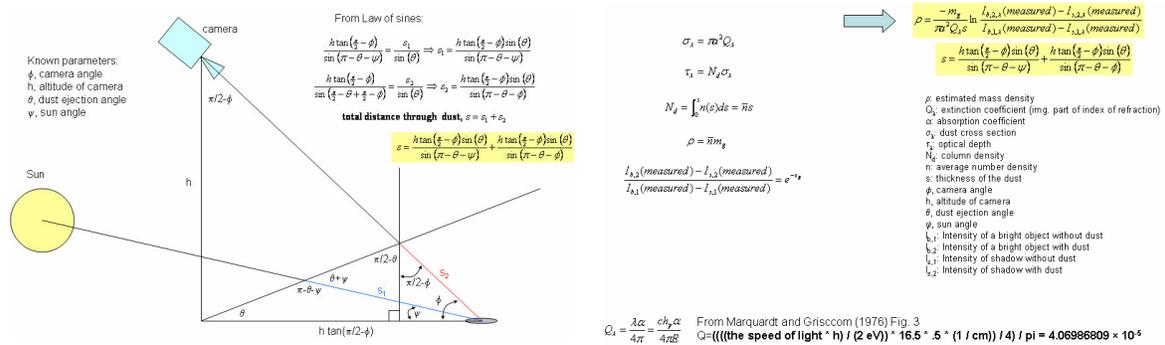

**Figure 6 Dust density calculations using data from the Apollo Landing Videos and properties of the lunar soil (Marquardt and Griscom 1976).**

The density of the lofted regolith has been estimated using the extinction coefficient of lunar regolith, the camera height, the sun angle, the camera angle, the dust ejection angle, and comparisons of a bright object during momentary clearings of the dust. Estimates for these events from Apollo 11 and 16 result in particle densities from $10^8$-$10^{13}$ particles/m$^3$: a few orders of magnitude larger than that predicted by Roberts' model.

*2.5  Rock Exhumation on Crater Rim*

The common belief during the Apollo program was that the plume impingement would result in excavation of regolith of size on the order of 1 millimeter and smaller. In fact, Conrad remarked on a exposed 3x4x2 inch rock that remained under the engine bell after the Apollo 12 landing (Wagner 2006).

For the Apollo 14 landing at about T-3.2 seconds (prior to touchdown), the plume exhumes two rocks on a crater rim in the central field of view. Shortly after, they are lofted by the exhaust plume. It is unclear from the videos whether the rocks are transported intact or whether they are disintegrated by the exhaust. Figure 7 shows four consecutive frames of the rock exhumation.

Using the altitude of the Lander, the magnification of the camera, and the effective pixel size the size of circular objects can be estimate via the calculations shown in Figure 8.

The authors estimate the rocks to be about 11-15 cm (4-6 inches) in diameter, much larger than that thought to be possible during the Apollo program. The Constellation program will need to address mitigation for moving objects of this size.

## 2.6 Streak Jumps From Thrust Change

The predominant theory for plume impingement onto regolith surfaces at the time of the Apollo program was developed by Leonard Roberts (Roberts 1963; Roberts 1966). The theory is an elegant analytical solution that accurately models many of the observed phenomena in an era predating numerical simulations. The theory does have some shortfalls. Roberts' model will be addressed in more detailed in future papers, however there are some insights that can be obtained from analysis of the Apollo landing videos (Lane, Metzger et al. 2008; Metzger, Lane et al. 2008).

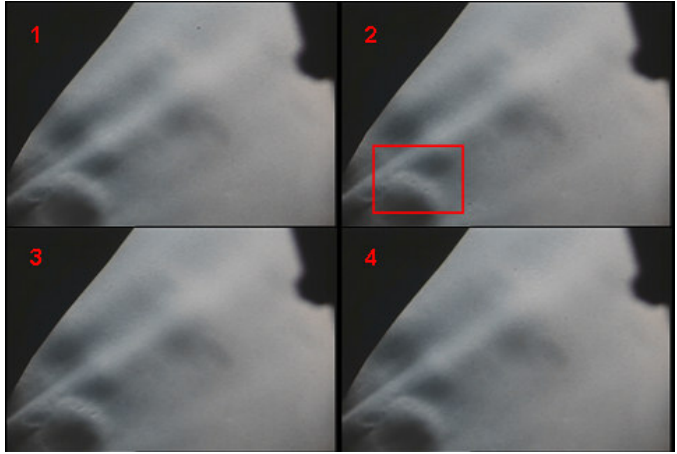

**Figure 7 Four sequential frames in the Apollo 14 landing video at about 3.2 seconds before touch-down. Note the rocks exhumed in frame 2 that are lofted away by the engine exhaust. The authors estimate the rocks at bout 11-15 cm diameter.**

In this model, the predominant mechanism for lofting of dust particles, to first order, is the wall angle of the inclined surface of the craters. Consequently, according to Roberts' theory, dust ejection angle is only weakly coupled to thrust level of the engine. While the "roughness" of the lunar surface certainly contributes to regolith excavation, there are several cases where ejection angle changes with engine thrust. Figure 9 shows an example of a "jump" in dust streaks between two consecutive frames. It is highly unlikely that the crater inner wall changes enough to correspond to the large change indust ejection angle between these two frames (~30 ms). This large change in ejection angle does correspond very closely in time with a change in Lander thrust.

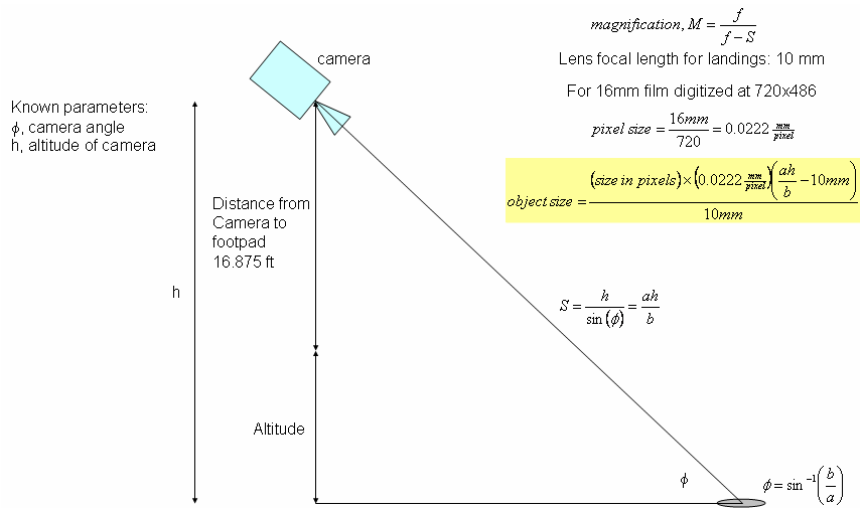

**Figure 8 Calculations for object size estimation from Apollo landing videos.**

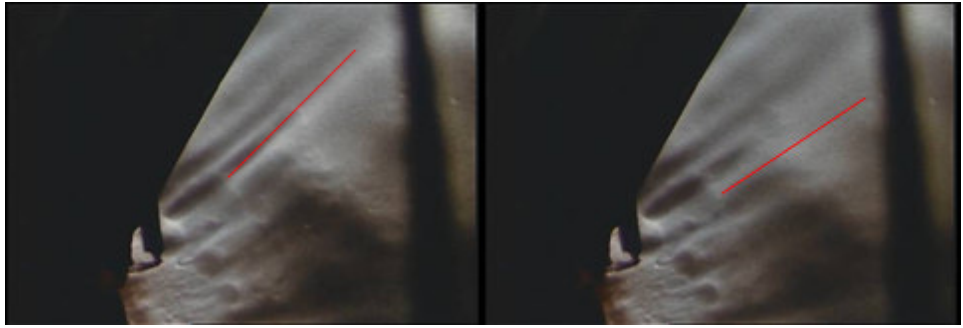

**Figure 9 Dust ejection angle change noted from a jump in the "streaks" between two sequential frames from the Apollo 12 landing video.**

## 3  Conclusions

For each of the six Apollo lunar landings, the descent videos have provided a wealth of information to the public. While most of the data to date have been qualitative in nature, it is possible to extract quantitative estimates from the videos. The authors estimate that the dust ejection angle derived from the exhaust plume was approximately 1-3 degrees on average, however it is possible to loft dust at much larger angles for certain, worst-case scenarios. From the landing videos, the estimate for dust loading in the exhaust is approximately $10^8$-$10^{15}$ particles/m$^3$. While a majority of the material eroded from beneath the landing site will be sub-millimeter sized regolith, it is possible that 10-15 cm sized objects can be lofted in the exhaust plume.

The current status of the modeling of plume/regolith interaction during landing is insufficient to quantitatively describe the complete physical scenario during landing. For the design of the landing sites for the Constellation program's return to the moon, the modeling must be improved to prevent damage to hardware at the landing sites. Additional effort will be put forth to quantify other phenomena in the landing videos to help plan for future missions to the moon.